# Gate-tunable Room-temperature Ferromagnetism in Two-dimensional Fe$_3$GeTe$_2$

Yujun Deng[1,2,3,†], Yijun Yu[1,2,3,†], Yichen Song[1,2,3], Jingzhao Zhang[4], Nai Zhou Wang[2,5,6], Yi Zheng Wu[1,2], Junyi Zhu[4], Jing Wang[1,2], Xian Hui Chen[2,5,6] and Yuanbo Zhang[1,2,3*]

[1]*State Key Laboratory of Surface Physics and Department of Physics, Fudan University, Shanghai 200438, China*

[2]*Collaborative Innovation Center of Advanced Microstructures, Nanjing University, Nanjing, 210093, China*

[3]*Institute for Nanoelectronic Devices and Quantum Computing, Fudan University, Shanghai 200433, China*

[4]*Department of Physics, the Chinese University of Hong Kong, Shatin, New Territories, Hong Kong, China*

[5]*Hefei National Laboratory for Physical Science at Microscale and Department of Physics, University of Science and Technology of China, Hefei, Anhui 230026, China*

[6]*Key Laboratory of Strongly Coupled Quantum Matter Physics, University of Science and Technology of China, Hefei, Anhui 230026, China*

† These authors contributed equally to this work.

* Email: zhyb@fudan.edu.cn

**Material research has been a major driving force in the development of modern nano-electronic devices. In particular, research in magnetic thin films has revolutionized the development of spintronic devices**[1–8]**; identifying new magnetic materials is key to better device performance and new device paradigm. The advent of two-dimensional van der Waals crystals creates new possibilities. This family of materials retain their chemical stability and structural integrity down to monolayers and, being atomically thin, are readily tuned by various kinds of gate modulation**[9–11]**. Recent experiments have demonstrated that it is possible to obtain two-dimensional ferromagnetic order in insulating $Cr_2Ge_2Te_6$ (ref. 12) and $CrI_3$ (ref. 13) at low temperatures. Here, we developed a new device fabrication technique, and successfully isolated monolayers from layered metallic magnet $Fe_3GeTe_2$ for magnetotransport study. We found that the itinerant ferromagnetism persists in $Fe_3GeTe_2$ down to monolayer with an out-of-plane magnetocrystalline anisotropy. The ferromagnetic transition temperature, $T_c$, is suppressed in pristine $Fe_3GeTe_2$ thin flakes. An ionic gate, however, dramatically raises the $T_c$ up to room temperature, significantly higher than the bulk $T_c$ of 205 Kelvin. The gate-tunable room-temperature ferromagnetism in two-dimensional $Fe_3GeTe_2$ opens up opportunities for potential voltage-controlled magnetoelectronics**[14,15]** based on atomically thin van der Waals crystals.**

Atomically thin, layered van der Waals (vdW) crystals represent ideal two-dimensional (2D) material systems with exceptional electronic structures. Since the discovery of graphene[16], such crystals have been at the frontier of material research, where vast opportunities arise from i) emerging physical properties as a result of reduced dimensionality, and ii) new gating capabilities to modulate the material properties now that the entire crystal is a surface. The former underpins the rapid permeation of 2D materials in areas ranging from semiconductors[17–19] to highly correlated material[11,20,21] and superconductor[22–24]. The trend continues with recent addition of magnetic crystals in the 2D material family[12,13]. In particular, intrinsic ferromagnetic order has been observed in monolayer $CrI_3$ in the form of a 2D Ising ferromagnet[13].

Here we report the discovery of 2D itinerant ferromagnetism in atomically thin $Fe_3GeTe_2$ (FGT) crystals. As in the case of $CrI_3$, intrinsic magnetocrystalline anisotropy in FGT monolayers counteracts thermal fluctuations and preserves the 2D long-rang ferromagnetic order, which is otherwise precluded in an isotropic magnetic system according to the Mermin-Wagner theorem[25]. FGT, however, offers a critical advantage: its metallic nature enables the interplay of both spin and charge degrees of freedom that lies at the heart of various spintronic architectures[26]. In this study we focus on the large anomalous Hall effect (AHE) resulting from such interplay. The AHE enables us to extract the ferromagnetic transition temperature $T_c$, and elucidate the evolution of the magnetic order from bulk down to monolayer FGT. Taking advantage of the tunability of atomically thin crystals, we further show that extreme doping induced by an ionic gate dramatically elevates the $T_c$ up to room temperature, accompanied by large modulations in the coercivity. These results establish FGT as a new itinerant ferromagnetic 2D material potentially suitable for electrically controlled magnetoelectronic devices operating at room temperature.

A variety of metallic layered compounds exhibit magnetism[27]. Among them FGT stands out with a relatively high $T_c$ (ref. 28, 29, ranging from 150 K to 220 K depending on Fe occupancy[30,31]). The atomic structure of a monolayer FGT is shown in Fig. 1b. In each of FGT monolayer, covalently-bonded $Fe_3Ge$ heterometallic slab is sandwiched between two Tellurium (Te) layers. The structure and valence states of the compound can be written as $(Te^{2-})(Fe_I^{3+})[(Fe_{II}^{2+})(Ge^{4-})](Fe_I^{3+})(Te^{2-})$ per formula with two inequivalent Fe sites, $Fe_I^{3+}$ and $Fe_{II}^{2+}$, within the $Fe_3Ge$ slab[29]. Partially filled Fe $d$ orbitals dominate the band structure around the Fermi level, and give rise to itinerant ferromagnetism in bulk FGT[32]. Adjacent monolayers are separated by a 2.95 Å van der Waals gap in the bulk crystal. As a result of the reduced crystal symmetry of the layered structure, bulk FGT exhibits a strong magnetocrystalline anisotropy[32,33]. Such anisotropy is expected to lift the restriction imposed by Mermin-Wagner theorem and stabilize the long-range ferromagnetic order in FGT monolayers.

Experimentally isolating atomically thin FGT crystals from the bulk, however, poses a challenge. Although bulk FGT cleaves along the vdW gap, the intra-layer bonding is not strong enough for thin flakes with reasonable size (e.g. 5 μm) to survive conventional mechanical exfoliation process. Similar issues plague the exfoliation of many other layered materials (only a small fraction of known layered crystals is cleavable down to monolayers), and have largely impeded the research on 2D materials in general.

To this end, we developed an $Al_2O_3$-assisted exfoliation method that enables isolation of monolayers from bulk layered crystals that are otherwise difficult to exfoliate with conventional method. This method was inspired by the gold-mediated exfoliation of layered transition-metal chalcogenide compounds[34,35]. The fabrication process is illustrated step-by-step in Fig. 1a. We started by covering the freshly cleaved surface of the bulk crystal with $Al_2O_3$ thin film with thickness ranging from 50 nm to 200 nm. The film was deposited by thermally evaporating Al under an oxygen partial pressure of $10^{-4}$ mBar. We then used a thermal release tape to pick up the $Al_2O_3$ film, along with pieces of FGT micro-crystals separated from the bulk. The $Al_2O_3$/FGT stack was subsequently transferred onto a piece of polydimethylsiloxane (PDMS), with the FGT side in contact with PDMS surface. Next, we quickly peeled away PDMS, leaving $Al_2O_3$ film covered with freshly cleaved FGT flakes on a substrate. Fig. 1c displays an optical image of atomically thin FGT flakes on $Al_2O_3$ film supported on a 285-nm $SiO_2$/Si substrate. Mono- to tri-layer regions of FGT are clearly resolved from the optical contrast. Such layer identification was corroborated by direct topography measurement with an atomic force microscope (AFM); the 0.8 nm steps in the height profile exactly match FGT monolayer thickness (Fig. 1d and e). We then fabricated electrodes on FGT thin flakes, either with direct metal deposition through stencil masks or with indium cold welding, for subsequent transport measurements. The entire device fabrication process was performed in an argon atmosphere with $O_2$ and $H_2O$ content kept below 0.5 ppm to avoid sample degradation. Details of the sample exfoliation and electrode fabrication are presented in Supplementary Section II. Finally, we point out that even though we only focus on FGT in this study, the method should be applicable to a wide variety of vdW crystals. We attribute the drastically enhanced exfoliation capability to higher affinity, as well as increased contact area, between evaporated $Al_2O_3$ film and freshly-cleaved FGT surface.

We study the magnetism in FGT by probing the Hall resistance, $R_{xy}$, under an external magnetic field, $\mu_0 H$, applied perpendicular to the vdW plane. In our samples that typically has a van der Pauw configuration, the fact that $R_{xy}$ is antisymmetric with respect to the polarity of $\mu_0 H$, whereas magnetoresistance is symmetric, enables us to separate the contribution of magnetoresistance from the raw Hall voltage data. For a magnetic material, $R_{xy}$ can be decomposed into two parts[36]:

$$R_{xy} = R_{NH} + R_{AH} \qquad (1)$$

where $R_{NH} = R_0 \mu_0 H$ is the normal Hall resistance, and $R_{AH} = R_S M$ is the anomalous Hall resistance; $R_0$ and $R_S$ are constant coefficients characterizing the strength of $R_{NH}$ and $R_{AH}$, respectively. Crucial information on the long-range magnetic order, which is contained in $M$, can therefore be extracted from the measurement of the anomalous Hall resistance.

Ferromagnetism persists in atomically thin FGT flakes down to monolayer. Unambiguous evidence comes from the clear hysteresis in $R_{xy}$ for all FGT thin flakes (with thickness ranging from monolayer to 50 layers in the bulk limit; Fig. 2a) under investigation at low temperatures. Such hysteresis reflects the hysteresis in $M$; the remnant $M$ at $\mu_0 H = 0$ signifies spontaneous magnetization, and thus the long-range ferromagnetic order, in two-dimensional FGT. Raising the temperature introduces thermal fluctuations to the ferromagnetic order, and ferromagnetism eventually disappears above $T_c$. In Fig. 2b, measurements of $R_{xy}$ show that ferromagnetism in samples with various layer numbers respond differently to thermal fluctuations: at an elevated temperature of $T = 100$ K hysteresis vanishes in mono- and bilayer samples, but survives in thicker crystals. The observation clearly indicates a strong dimensionality effect on ferromagnetism in atomically thin FGT.

To fully elucidate the effect of reduced dimensionality on ferromagnetism in 2D FGT, we precisely determine $T_c$ as a function of the number of layers. For each sample thickness, we examine the remnant Hall resistance at zero external magnetic field, $R_{xy}^r \equiv R_{xy}|_{\mu_0 H=0}$, as a function of temperature. According to equation (1), $R_{xy}^r$ is directly proportional to the zero-field spontaneous magnetization $M|_{\mu_0 H=0}$. So the onset of nonzero $R_{xy}^r$ indicates the emergence of spontaneous magnetization. The temperature at the onset therefore marks the $T_c$, where long-range ferromagnetic ordering start to set in (Fig. 2c; ref. 37). Analysis for various sample thicknesses reveals that $T_c$ decreases monotonically as the samples are thinned down, from ~ 200 K in the bulk limit to 20 K in a monolayer. The one-order-of-magnitude decrease in $T_c$ implies a dramatic reduction in the energy scale of magnetic ordering in 2D FGT, which we shall discuss next.

The strong dimensionality effect on the ferromagnetism in atomically thin FGT stems from the fundamental role of thermal fluctuation in 2D. The itinerant ferromagnetism in FGT is, in principle, described by Stoner model for metallic magnetic systems[29,32,33]. It has, however, been established that such itinerant ferromagnetism can be mapped on to a classical Heisenberg model with Ruderman-

Page 5 of 17

Kittel-Kasuya-Yosida (RKKY) exchange[38]. Armed with this insight, we adopt the following anisotropic Heisenberg Hamiltonian in the absence of external magnetic field:

$$H = \sum_{i,j} J_{ij} \mathbf{S}_i \cdot \mathbf{S}_j + \sum_i A(S_i^z)^2 \qquad (2)$$

where $\mathbf{S}_i$ is the spin operator on site $i$, and $J_{ij}$ the exchange coupling between spins on site $i$ and $j$; $A$ is the single-ion perpendicular magnetocrystalline anisotropy arising from spin-orbit coupling. In the three-dimensional (3D) bulk limit, the density of states per spin for the magnon modes is strongly suppressed, so thermal fluctuations destroy the long-range magnetic order only at a finite $T_c$ determined primarily by exchange interactions $J_{ij}$ (ref. 12). The bulk $T_c$ of ~ 200 K implies that the energy scale of $J_{ij}$ (summed over nearest neighbours) in FGT is on the order of ~ 10 meV. In monolayer FGT, however, the largely isotropic exchange couplings $J_{ij}$ (see Supplementary Section IV for detailed *ab initio* calculations) alone will not be able to sustain magnetic order at finite temperatures because of thermal fluctuations of the long-wavelength gapless magnon modes in 2D. In this case, the magnetocrystalline anisotropy $A$ gives rise to an energy gap in the magnon dispersion. The gap suppresses low-frequency, long-wavelength magnon excitations, and protects the magnetic order below a finite $T_c$ determined primarily by $A$. The monolayer $T_c$ of 20 K implies an $A$ on the order of 0.6 meV. The estimations of both exchange interactions and magnetocrystalline anisotropy are in reasonable agreement with results from *ab initio* density functional calculations (ref. 32, 39; see also Supplementary Section IV). The magnetocrystalline anisotropy corresponds to an energy density $K_u$ on the order of $10^6$ Jm$^{-3}$, which is one order of magnitude larger than interfacial magnetic anisotropy in a 1.3-nm-thick CoFeB film[40]. Such a large intrinsic perpendicular anisotropy is important for potential stable, compact spintronic devices based on atomically thin FGT.

The evolution from 3D to 2D magnetism in FGT is linked with the critical behaviour at the paramagnet (PM) to ferromagnet (FM) phase transition. Because $T_c$ is the critical temperature where spin-spin correlation length diverges, a finite sample thickness limits the divergence of the correlation length, and therefore depresses $T_c$. Analyses of the critical behaviour[41–43] show that layer-number-dependent $T_c(N)$ follows a universal scaling law as the sample thickness approaches the 2D limit: $[T_c(\infty) - T_c(N)]/T_c(\infty) = ((N_0 + 1)/2N)^\lambda$, where $T_c(\infty)$ denotes $T_c$ of the bulk crystal. The critical exponent $\lambda = 1/\nu$ reflects the universality class of the transition; $N_0$ is the critical layer number defined by the mean spin-spin interaction range. Analyses of thin ferromagnetic metal films[43–46] also reveals that the power law is replaced by a linear relation when the film is thinner than the critical value, i.e. $N \leq N_0$. $N_0$, therefore, marks the boundary that separates 2D magnetism and 3D magnetism in ultra-thin ferromagnets. The scaling law fits well to the $T_c(N)$ measured in FGT as shown in Fig. 2d (here we fitted the highest $T_c$ achieved for each sample thickness). The fit yields a critical layer number of $N_0 = 3.2 \pm 0.6$ and a critical exponent of $\lambda = 1.7 \pm 0.6$. We also confirmed that the linear relation well describes our data in the thinnest samples ($N \leq 3$; see Fig. 2d). We note that $N_0$ obtained in FGT is comparable to that in the magnetic thin films[43,44,47]. Meanwhile, $\lambda$ lies between the values expected from mean field model ($\lambda = 2$) and 3D Ising model ($\lambda = 1.587$), although it's also consistent with 3D Heisenberg model ($\lambda = 1.414$); more accurate determination of $T_c$ is required to resolve the universality class of bulk FGT.

Even though $T_c$ is suppressed in atomically thin FGT, we now demonstrate that an ionic gate could drastically modulate the ferromagnetism in FGT thin flakes, and boost the $T_c$ up to room temperature. We employ an ionic field-effect transistor setup, in which solid electrolyte (LiClO$_4$ dissolved in polyethylene oxide (PEO) matrix) covers both the FGT thin flake and a side gate (ref 11; Fig. 3a inset). A positive gate voltage, $V_g$, intercalates lithium ions into the FGT thin flake, much like the charging process in a lithium-ion battery. The charge transfer between lithium ions and the host crystal induces electron doping up to the order of $10^{14}$cm$^{-2}$ per layer[11]. The unprecedented doping level effects profound change in the ferromagnetism in FGT thin flake. Fig. 3b and 3c display the $R_{xy}$ of a tri-layer sample obtained at $T = 10$ K and $T = 240$ K, respectively. A moderate $V_g$ of a few volts induces significant variations in the coercivity of the ferromagnetic FGT, and the ferromagnetism, manifested as a finite $R_{xy}^r$, persists at high temperatures under certain gate doping. Indeed, detailed analysis reveals that $T_c$ is dramatically modulated by the ionic gate with the highest $T_c$ reaching room temperature ($T = 300$ K; Fig. 3d). The gate-induced room-temperature ferromagnetism is also directly observed in a four-layer sample as shown in Fig. 4. $T_c$ as a function of the gate voltage exhibits a complex pattern: after an initial drop, $T_c$ increases sharply to room temperature at $V_g = 1.75$ V. $T_c$ peaks again at $V_g = 2.39$ V on top of an otherwise slowly declining trend as the doping level increases (Fig. 3d; see Supplementary Section III for details). The variations in $T_c$ is accompanied by changes in the sample conductance, $\sigma_{xx}$, obtained at $T = 330$ K (Fig. 3a). In particular, the two peaks in $T_c$ coincide with sudden increases in $\sigma_{xx}$, suggesting changes in the electronic structure as the origin of the magnetism modulation. Finally, we note that the coercivity measured at $T = 10$ K also roughly follows the variations in $T_c$, as shown in Fig. 3e.

The gate-tuned ferromagnetism in atomically thin FGT is consistent with Stoner model that describes itinerant magnetic systems. The Stoner criterion infers that the formation of ferromagnetic order is dictated by the density of states (DOS) at the Fermi level[48]. The extreme electron doping induced by the ionic gate causes substantial shift of FGT's electronic bands. Consequently, the large variation in the DOS at the Fermi level leads to appreciable modulation in the ferromagnetism. Detailed calculations support such a picture. Specifically, our calculation shows that gate-induced electrons sequentially fill the sub-bands originated from Fe $d_{z^2}$, $d_{xz}$ and $d_{yz}$ orbitals, leading to sharp peaks in DOS that corresponds to Fermi level's passing through the flat band edges of the sub-bands (see Supplementary Section IV). Such DOS peaks may be responsible for the sharp increase of $T_c$ and the corresponding sudden jump in $\sigma_{xx}$. Meanwhile, calculations indicate that change in the average magnet moment per Fe atom is minimal (< 3%) under charge doping (Supplementary Section IV). This result suggests that the gate-tuned coercivity shown in Fig. 3e may in fact reflect the large modulation in the anisotropy energy, although we cannot rule out effects from domain nucleation that complicates the correlation between coercivity and anisotropy.

To conclude, we develop a new fabrication method and successfully isolated atomically thin FGT from the layered bulk crystal. Itinerant ferromagnetism persists in 2D FGT down to monolayer. The ferromagnetism is protected by a perpendicular magnetocrystalline anisotropy against thermal fluctuations in the 2D limit, and exhibits strong dimensionality effect as the thickness increases. The atomically thin FGT provides a rare opportunity for electrical modulation of its magnetic properties. We achieve unprecedented control of its ferromagnetism with an ionic gate. In particular, we realize room-temperature ferromagnetic order in tri-layer and four-layer FGT under gate modulation. Our discovery of ferromagnetic FGT provides an ideal model system for studying 2D itinerant ferromagnetism. The high $T_c$ achieved in FGT thin crystals, combined with the large perpendicular anisotropy, will be important for the room-temperature operation of potential ultra-high density, gate-tunable magnetoelectronic devices based on 2D FGT.

49. Ueno, K. *et al.* Anomalous Hall effect in anatase Ti1−xCoxO2−δ above room temperature. *J. Appl. Phys.* **103,** 07D114 (2008).

**Acknowledgements**

We thank Xiaofeng Jin, Biao Lian and Gang Chen for helpful discussions. Part of the sample fabrication was conducted at Fudan Nano-fabrication Laboratory. D.Y., Y.Y., S.Y., J.W. and Y.Z. acknowledge financial support from National Key Research Program of China (grant no. 2016YFA0300703), and NSF of China (grant nos. U1732274, 11527805 and 11425415). N.Z.W. and X.H.C. acknowledge support from the 'Strategic Priority Research Program' of the Chinese Academy of Sciences (grant no. XDB04040100) and the National Basic Research Program of China (973 Program; grant no. 2012CB922002). X.H.C. also acknowledges support from the National Natural Science Foundation of China (grant no. 11534010) and the Key Research Program of Frontier Sciences, CAS (grant No. QYZDY-SSW-SLH021). J.Y.Z acknowledge financial support from Chinese University of Hong Kong (CUHK) under Grant No 4053084, from University Grants Committee of Hong Kong under Grant No 24300814, and the Start-up Funding of CUHK. J.W. also acknowledges support from the NSF of China (grant no. 11774065).

**Author contributions**

Y.Z. conceived the project. N.Z.W. and X.H.C. grew bulk FGT crystal. Y.D., Y.S. and Y.Y. developed device fabrication method. Y.D. fabricated devices, and performed electric measurements with the help of Y.Y.. Y.D., Y.Y. and Y.Z. analyzed the data. J.Z.Z and J.Y.Z carried out DFT calculations; J.W. carried out theoretical calculations and modeling; and all three performed theoretical analysis. Y.D., Y.Y., J.W. and Y.Z. wrote the paper and all authors commented on it.

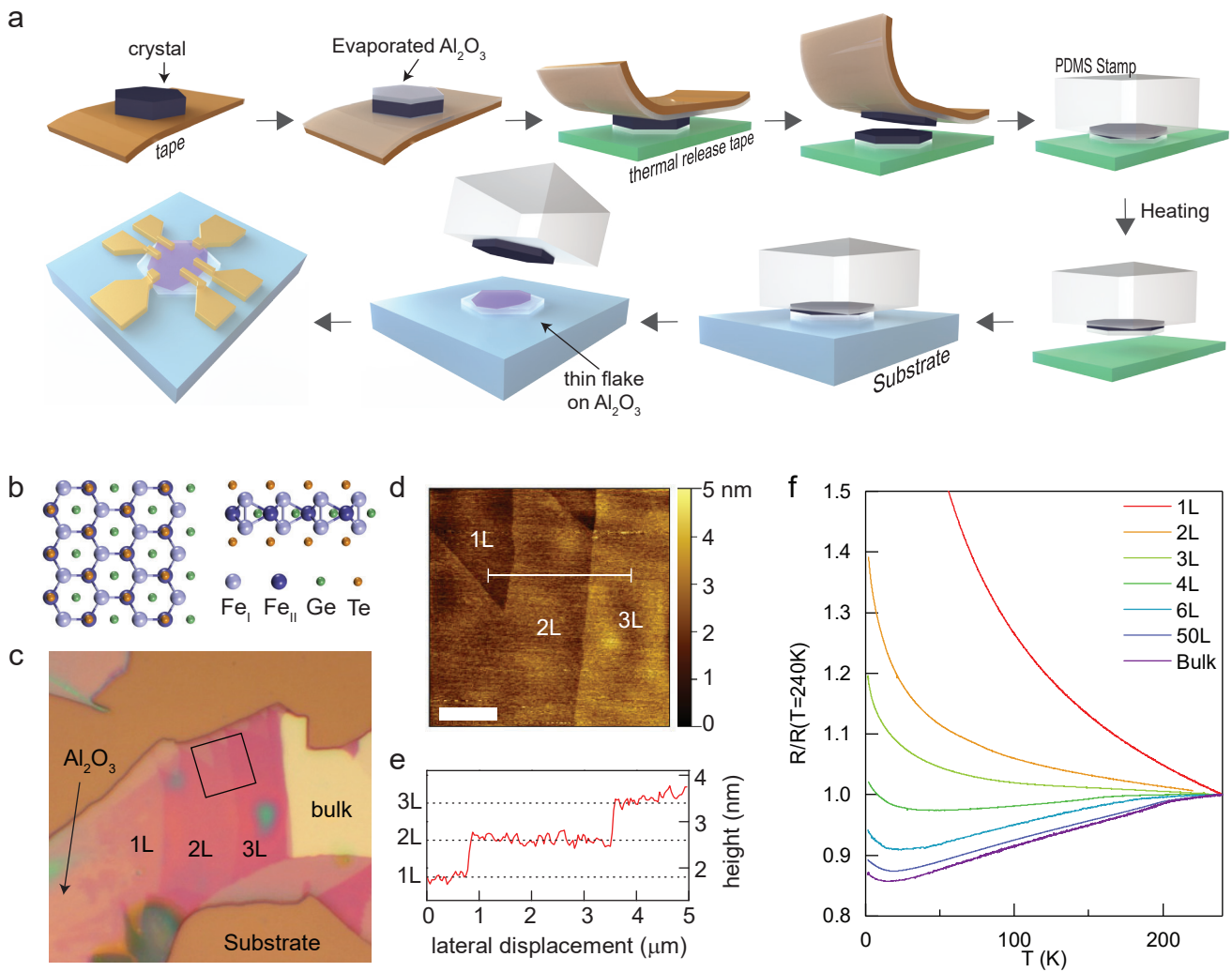

**Figure 1 | Fabrication and characterization of atomically-thin FGT devices. a**, Schematic of a new $Al_2O_3$-assisted mechanical exfoliation method. The new method uses an $Al_2O_3$ film evaporated onto the bulk crystal to cleave layered materials. The strong adhesion between crystal and $Al_2O_3$ film makes it possible to exfoliate layered crystals that are otherwise difficult to cleave with $SiO_2$ surface in conventional methods. **b**, Atomic structure of monolayer FGT. Left: view along [001]; right: view along [010]. Bulk FGT is a layered crystal with interlayer vdW gap of 2.95 Å (ref. 29). $Fe_I$ and $Fe_{II}$ represent the two inequivalent Fe sites in +3 and +2 oxidation state, respectively. **c**, Optical image of typical few-layer FGT flakes exfoliated on top of $Al_2O_3$ thin film. The $Al_2O_3$ film is supported on Si wafer covered with 285 nm $SiO_2$. **d**, AFM image of the area marked by the square in **c**. Mono- and few-layer flakes of FGT are clearly visible. Scale bar: 2 μm. **e**, Cross-sectional profile of the FGT flakes along the white line shown in **d**. The steps are 0.8 nm in height, consistent with the thickness of monolayer FGT (0.8 nm). **f**, Temperature-dependent sample resistance of FGT with varying number of layers. Resistances are normalized to their values at T=240 K. Typical behaviour of bulk FGT (purple) is also shown here for reference.

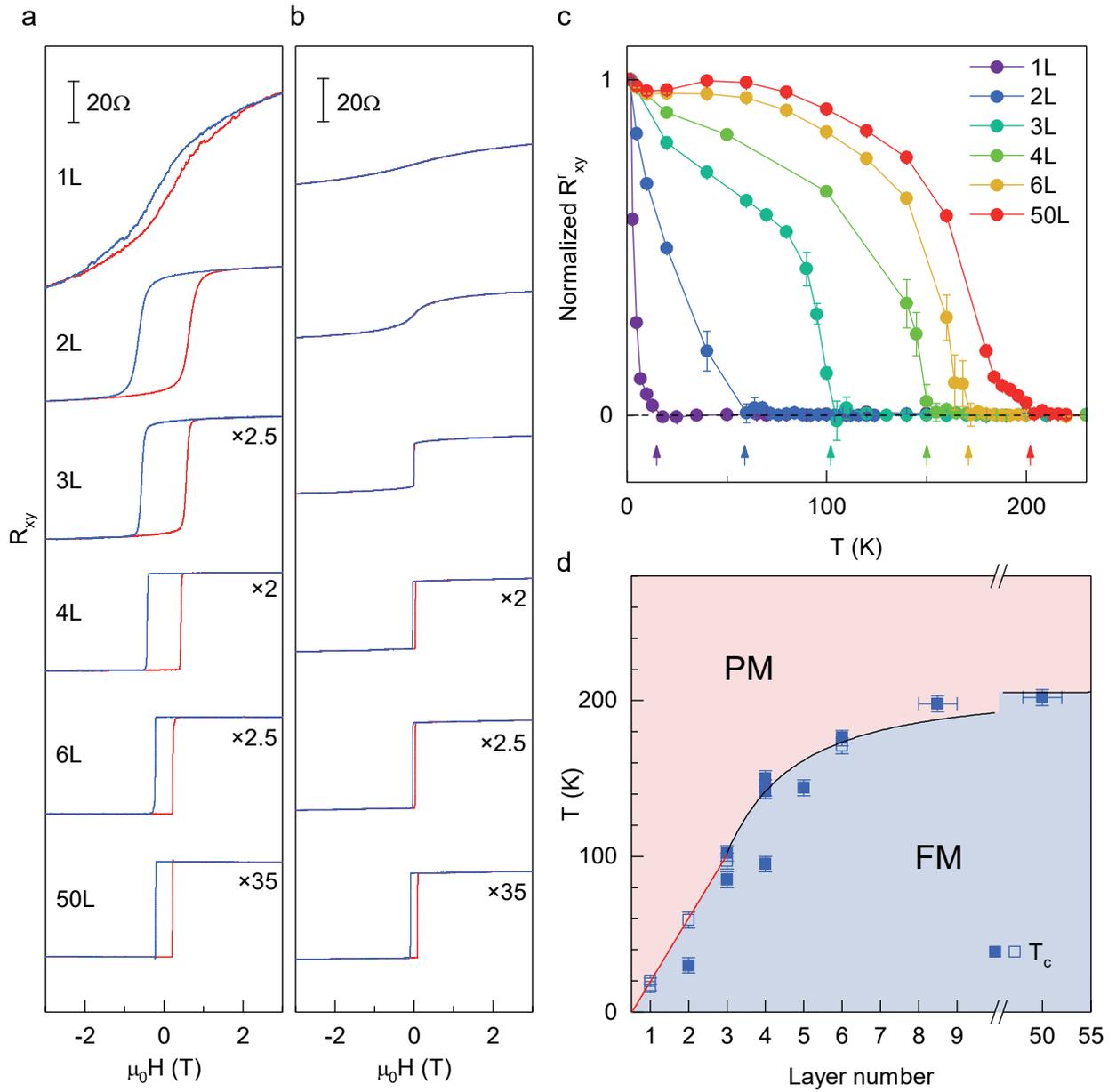

**Figure 2 | Ferromagnetism in atomically-thin FGT. a,b**, $R_{xy}$ at low temperature (**a**) and 100 K (**b**) obtained in FGT thin-flake samples with varying number of layers. Ferromagnetism, evidenced by the hysteresis in low-temperature $R_{xy}$, is observed in all atomically-thin FGT down to monolayer. At $T = 100$ K, hysteresis disappears in mono- and bilayer FGT, but persists in thicker specimens. All the low temperature data were obtained at $T = 1.5$ K except for monolayer data, which were recorded at $T = 3$ K. **c**, Remnant anomalous Hall resistance $R_{xy}^r$ as a function of temperature obtained from FGT thin-flake samples with varying number of layers. $R_{xy}^r$ are normalized to their values at $T = 1.5$ K except for the monolayer dataset, which is normalized to the $T = 3$ K value. Arrows mark the ferromagnetic transition temperature $T_c$. **d**, Phase diagram of FGT as layer number and temperature are varied. Filled squares represent $T_c$ of samples with Cr/Au electrodes made by direct evaporation through stencil masks, and open squares represent $T_c$ of samples with cold-welded indium microelectrodes. Back line is a fit to $T_c(N)$ (for $N \geq 3$) by the finite-size scaling formula described in the main text. Solid red line is line fit to $T_c(N)$ for $N \leq 3$. Here we fitted the highest $T_c$ achieved for each sample thickness.

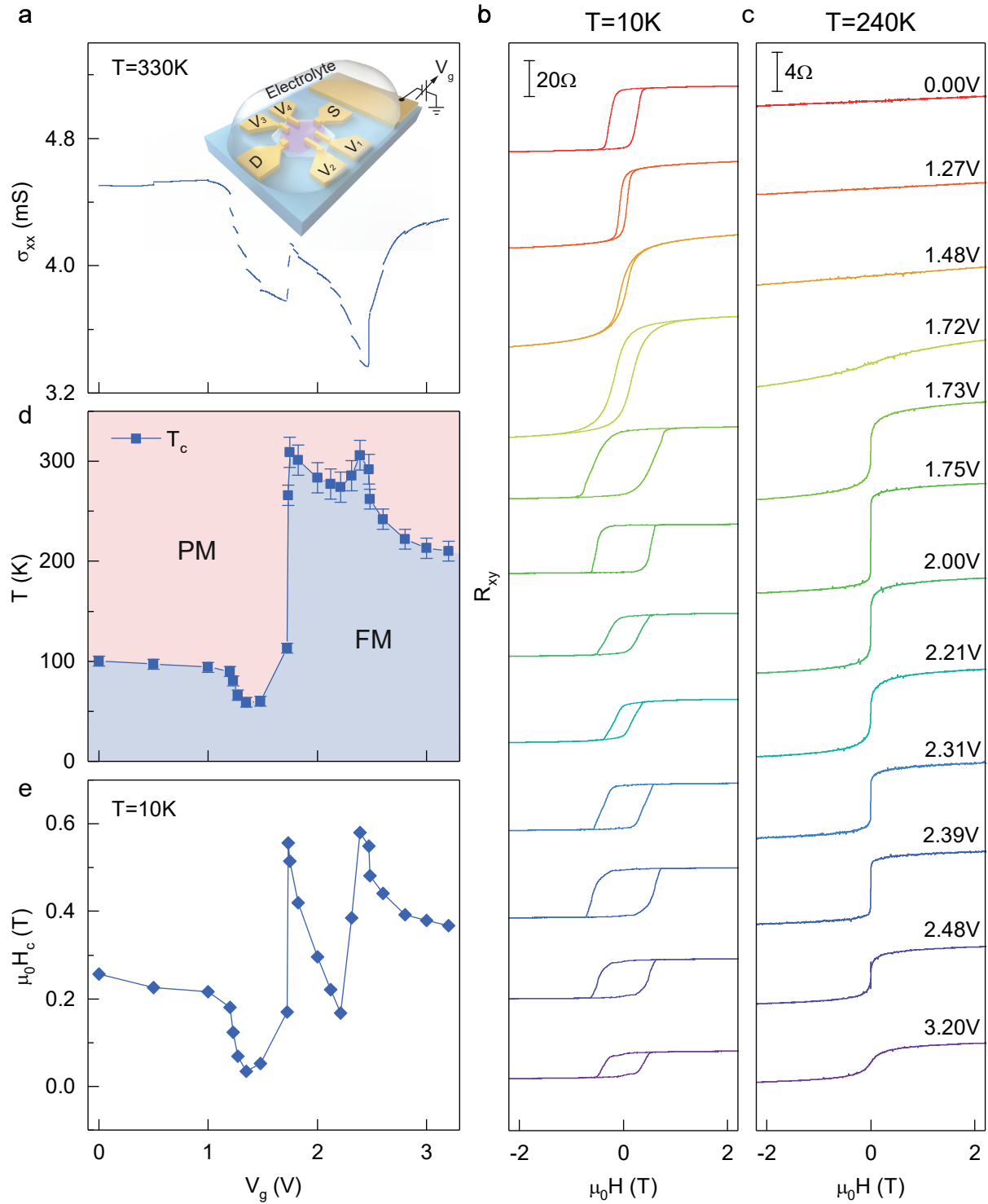

**Figure 3 | Ferromagnetism in atomically-thin FGT flake modulated by ionic gate. a**, Conductance as a function of gate voltage, $V_g$, measured in a tri-layer FGT device. Data were obtained at $T$ = 330 K. Breaks in the curve were caused by temperature-dependent measurements at fixed gate voltages during the gate-sweep. Inset: schematic of the FGT device structure and measurement setup. S and D label source and drain electrodes, respectively, and $V_1$, $V_2$, $V_3$ and $V_4$ label the voltage probes. Solid electrolyte LiClO$_4$/PEO covers both the FGT flake and the side gate. **b,c**, $R_{xy}$ as a function of external magnetic field recorded at representative gate voltages. Data obtained at $T$ = 10 K and $T$ = 240 K are displayed in **b** and **c**, respectively. **d**, Phase diagram of the tri-layer FGT sample as the gate voltage and temperature are varied. We determine the transition temperature from temperature-dependent anomalous Hall resistance extrapolates to zero (see Supplementary Section III). **e**, Coercive field as a function of the gate voltage. Data were obtained at $T$ = 10 K.

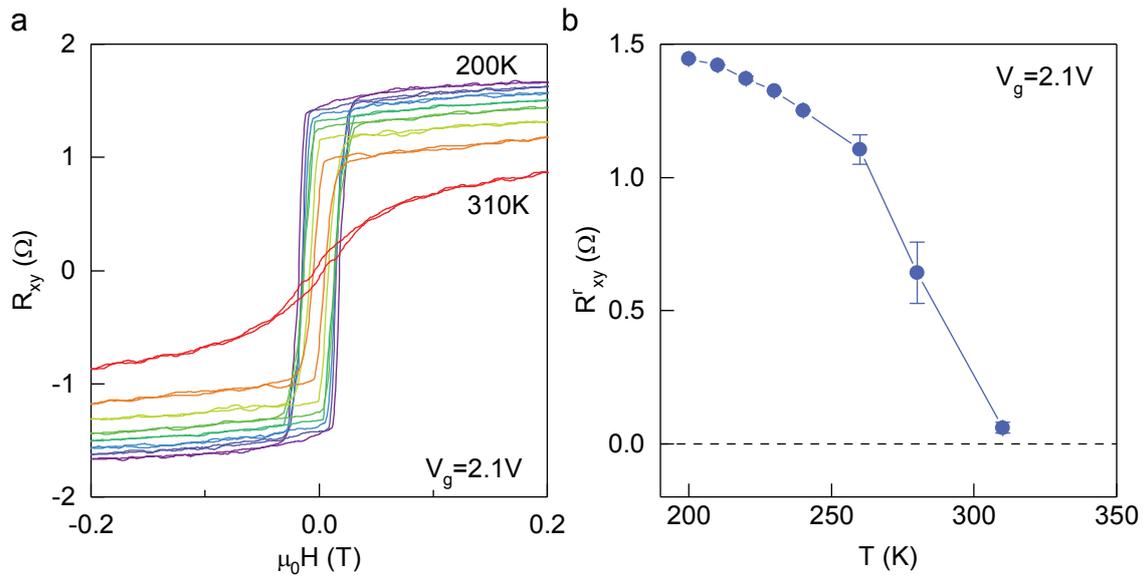

**Figure 4 | Direct observation of room-temperature ferromagnetism in atomically thin FGT. a**, $R_{xy}$ of a four-layer FGT under a gate voltage of $V_g = 2.1$ V. We have subtracted a background in $R_{xy}$ resulting from Lithium ions' slow drifting in the electrolyte under the gate bias above $T = 250$ K (ref. 49). Hysteresis in $R_{xy}$ persists up to $T = 310$ K, providing unambiguous evidence for room-temperature ferromagnetism in the sample. **b**, Remnant Hall resistance $R^r_{xy}$ as a function of temperature. Extrapolating $R^r_{xy}$ to zero yields a $T_c$ higher than 310 K.